# Gravity Gradient Tensor Eigendecomposition for Spacecraft Positioning

Pei Chen[1], Xiucong Sun[2], and Chao Han[3]
*Beihang University, Beijing, 100191, People's Republic of China*

## Nomenclature

| | | |
|---|---|---|
| $U$ | = | gravitational potential, m²/s² |
| **T** | = | gravity gradient tensor (GGT), E |
| **T**$_{ECEF}$ | = | GGT in ECEF, E |
| **T**$_{ENU}$ | = | GGT in ENU, E |
| **T**$_{EIG}$ | = | GGT in the Eigen frame, E |
| **T**$_0$ | = | GGT in GRF, E |
| **L**$_{A \to B}$ | = | rotation matrix from A to B |
| **r** | = | observation position, m |
| $x, y, z$ | = | cartesian coordinates of **r** |
| $r, \varphi, \lambda$ | = | spherical coordinates of **r** |
| **Q** | = | rotation matrix from GRF to inertial frame |
| **W** | = | rotation matrix from inertial frame to ECEF |
| $\mu$ | = | geocentric gravitational constant, m³/s² |
| $J_2$ | = | unnormalized second zonal harmonic |
| $R$ | = | equatorial radius of the Earth, m |
| $\xi$, **η** | = | eigenvalue and eigenvector |
| $\theta$ | = | rotation angle from ENU to the Eigen frame, rad |

## I. Introduction

GRAVITY gradient tensor (GGT) is an ideal observable for passive navigation where GNSS is undesirable or unavailable. Over the past few decades, a significant advance in gravity gradient instrument (GGI) technology has occurred due to geophysical exploration activities [1]. The electrostatic gravity gradiometer built by Thales Alenia Space, for example, demonstrated a noise density level of 0.01 E/√Hz within measurement bandwidth, and was successfully implemented on ESA's GOCE satellite [2]. A convert navigation system could be formed if data collected from this device are matched with a map of the Earth's gravity field.

---

[1] Associate Professor, School of Astronautics, cjhien@gmail.com.
[2] Doctoral Student, School of Astronautics, sunxiucong@gmail.com.
[3] Professor, School of Astronautics, hanchao@buaa.edu.cn.

Many researchers have studied the application of gravity gradiometry in navigation [3-9]. A general use is to incorporate gravity gradient measurements into a Kalman filter to prevent error accumulation of an inertial navigation system. Richeson showed that a GGI with noise level of 0.1 E could limit position error to roughly 10-20 meters under the assumption that the gradient maps were true [8]. Markley and Psiaki proposed an idea to simultaneously determine the orbits of two spacecraft from relative position vector measurements [10-11]. The two-satellite system was regarded as a very long baseline gravity gradiometer and its observability was investigated.

In this Note, a new approach to spacecraft positioning based on GGT inversion is presented. The gravity gradient tensor is initially measured in the gradiometer reference frame (GRF) and then transformed to the Earth-Centered Earth-Fixed (ECEF) frame via attitude information as well as Earth rotation parameters. Matrix Eigen-Decomposition is introduced to directly translate GGT into position based on the fact that the eigenvalues and eigenvectors of GGT specific functions of spherical coordinates of the observation position. . Unlike the strategy of inertial navigation aiding, no prediction or first guess of the spacecraft position is needed. The method makes use of the $J_2$ gravity model, and is suitable for space navigation where higher frequency terrain contributions to the GGT signals can be neglected.

## II. GGT Inversion and System Observability

The gravity gradient tensor **T** is the second-order gradient of the gravitational potential $U$, and is given by:

$$\mathbf{T} = \nabla(\nabla U) = \begin{bmatrix} \frac{\partial^2 U}{\partial x^2} & \frac{\partial^2 U}{\partial x \partial y} & \frac{\partial^2 U}{\partial x \partial z} \\ \frac{\partial^2 U}{\partial y \partial x} & \frac{\partial^2 U}{\partial y^2} & \frac{\partial^2 U}{\partial y \partial z} \\ \frac{\partial^2 U}{\partial z \partial x} & \frac{\partial^2 U}{\partial z \partial y} & \frac{\partial^2 U}{\partial z^2} \end{bmatrix} \qquad (1)$$

where $\nabla$ denotes the gradient operator, and $x$, $y$ and $z$ are the cartesian coordinates of position **r**. $U$ and **T** are scalar and matrix-valued functions of **r** respectively.

A full-tensor gradiometer measures gravity gradients along baselines of its three pairs of accelerometers, and GGTs are output in the gradiometer reference frame. Raw GGTs can be transformed to the ECEF frame via precise attitude information and Earth Orientation Parameters (EOP):

$$\mathbf{T}_{ECEF} = \mathbf{W} \cdot \mathbf{Q} \cdot \mathbf{T}_0 \cdot \mathbf{Q}^T \cdot \mathbf{W}^T \quad (2)$$

where $\mathbf{T}_0$ is GGT in GRF, $\mathbf{T}_{ECEF}$ is GGT in ECEF, $\mathbf{Q}$ represents the rotation matrix from GRF to the inertial frame, and $\mathbf{W}$ represents the rotation matrix from the inertial frame to ECEF. $\mathbf{Q}$ and $\mathbf{W}$ are related to attitude and EOP respectively.

After GGT transformation, positions can be extracted using a gravity model. This process is named as GGT inversion in this Note and is written as:

$$\hat{\mathbf{r}} = \mathbf{F}(\mathbf{T}_{ECEF}) \quad (3)$$

where $\hat{\mathbf{r}}$ is the resolved position, and the vector function $\mathbf{F}$ represents the operation of GGT inversion. A simplified block diagram of the prototype navigation system is shown in Fig. 1. Different with the gravity gradient map matching technique in [8], the system has no prediction-correction mechanism and can fix position on single-epoch, i.e., instantaneously.

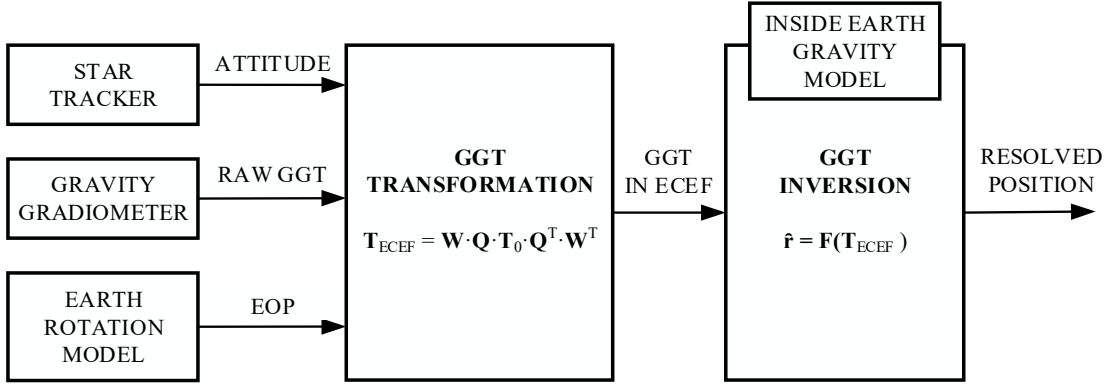

Fig. 1 Block Diagram of the GGT Inversion Positioning System.

A. Observability

By the continuity of the gravitational field, GGT is symmetric; and by Laplace's equation its trace is zero [12]. The nine elements of $\mathbf{T}_{ECEF}$ reduce to five independent terms. With three unknown position components and five gravity gradient observation components, an overdetermined nonlinear system is formulated. The Gramian matrix is a powerful tool for evaluating the system's observability as well as the covariance of position estimations [13], whereas this Note tries to explain the observability from geometry and give an intuitive method for error analysis.

To illustrate observability from geometry, the central gravity field where the Earth is assumed to be a point mass is used. The gravitational potential of the central gravity model is written as:

$$U = \frac{\mu}{r} \qquad (4)$$

where $\mu$ is the geocentric gravitational constant of the Earth, and $r$ is the distance from the Earth center to the observation point. GGT in ECEF is given by:

$$\mathbf{T}_{ECEF} = \begin{bmatrix} T_{xx} & T_{xy} & T_{xz} \\ T_{yx} & T_{yy} & T_{yz} \\ T_{zx} & T_{zy} & T_{zz} \end{bmatrix} = \frac{\mu}{r^3} \begin{bmatrix} -1+\frac{3x^2}{r^2} & \frac{3xy}{r^2} & \frac{3xz}{r^2} \\ \frac{3xy}{r^2} & -1+\frac{3y^2}{r^2} & \frac{3yz}{r^2} \\ \frac{3xz}{r^2} & \frac{3yz}{r^2} & -1+\frac{3z^2}{r^2} \end{bmatrix} \qquad (5)$$

From Eq. (5), the common factor, $\mu/r^3$, which can be viewed as the magnitude of GGT, is a strict monotonic function of $r$, and contains altitude information. The remaining dimensionless matrix is independent of $r$ and contains latitude and longitude information. Its three diagonal elements are positively related to the three components of the unit vector of $\mathbf{r}$, and the ratios between the off-diagonal elements are related to ratios between the position components.

An ambiguity of sign, however, exists in the dimensionless matrix. Locations at symmetrically opposite positions have the same value of $\mathbf{T}_{ECEF}$. The sign ambiguity causes a $2r$ difference between the two position solutions. The ambiguity problem is not studied in this Note. It is assumed that a prior position or a second navigation sensor kicks out the false position.

Another aspect of navigation observability is sensitivity, which measures the change of observations caused by state variations. High sensitivity is desired for all the navigation systems since any minor state error would be magnified and observed. For nonlinear systems, sensitivity is usually evaluated by linearization of the observation equations and can be measured by the magnitude of partial derivatives, which are called sensitivity factors here. The derivative of $\mu/r^3$ with respect to $r$ is $-3\mu/r^4$. Thus the sensitivity factor for the vertical position is $3\mu/r^4$. The value is about $7\times10^{-4}$ E/m near the Earth's surface, which means that a vertical position variation of 1000 m causes a $\mathbf{T}_{ECEF}$ magnitude variation of 0.7 E.

The sensitivity analysis is somewhat complicated for latitude and longitude. Taking the $T_{xx}$ component for example, the observation equation is:

$$T_{xx} = \frac{\mu}{r^3}\left(-1 + 3n_x^2\right) \quad (6)$$

where $n_x$ is the x component of the unit vector of **r**. The partial derivative of $T_{xx}$ with respect to $n_x$ is $6n_x\mu/r^3$. As $n_x$ varies between -1 and 1, the sensitivity factor has a maximum value of $6\mu/r^3$. The value is about $9\times10^3$ E near the Earth's surface, which means that a latitude or longitude variation of $10^{-4}$ rad causes a $T_{xx}$ variation of 0.9 E. The horizontal position is related to latitude and longitude by a coefficient of $r$. Thus the sensitivity factor for the horizontal position is $6\mu/r^4$, which is on the same order as that for the vertical position.

In spite of low sensitivity, GGT is attractive to navigation if the measurement noise level is low enough. For spacecraft in low earth orbits (LEO), the observability is acceptable. It is possible to design such an autonomous passive navigation system or use it as a back-up.

**B. Error Analysis for GGT inversion**

The position error of the GGT inversion system is attributed to errors in the gravity model and $\mathbf{T}_{ECEF}$ observations. Let $\delta\mathbf{T}_{ECEF}$ and $\Delta\mathbf{T}_{ECEF}$ denote the GGT model error and observation error respectively. Using the sensitivity factors given above, the vertical and horizontal position errors can be estimated by:

$$\begin{aligned}\Delta P_v &= \frac{r^4}{3\mu}\left(\|\delta\mathbf{T}_{ECEF}\|_{max} + \|\Delta\mathbf{T}_{ECEF}\|_{max}\right) \\ \Delta P_h &= \frac{r^4}{6\mu}\left(\|\delta\mathbf{T}_{ECEF}\|_{max} + \|\Delta\mathbf{T}_{ECEF}\|_{max}\right)\end{aligned} \quad (7)$$

where $\|\mathbf{A}\|_{max} \equiv \max\{|a_{ij}|\}$, representing the maximum norm of the elements.

The gravitational potential is usually represented as an infinite harmonic summation, which in spherical coordinates is expressed as:

$$U(r,\varphi,\lambda) = \frac{\mu}{r}\left[1 + \sum_{n=2}^{\infty}\left(\frac{R}{r}\right)^n \sum_{m=0}^{n}\left(\overline{C}_{nm}\cos m\lambda + \overline{S}_{nm}\sin m\lambda\right)\overline{P}_{nm}(\sin\varphi)\right] \quad (8)$$

where $R$ is the Earth's radius, $\varphi$ is latitude, $\lambda$ is longitude, $n$ and $m$ are degree and order, $\overline{C}_{nm}$ and $\overline{S}_{nm}$ are the normalized spherical harmonic coefficients, and $\overline{P}_{nm}$ is the associated normalized Legendre function. Terrain elevation contribution is negligible at sufficiently high altitudes. Referring to [8], a satellite in a 300 km altitude orbit with a space-grade 0.01 E GGI noise level would only be affected by terrain effects greater than 500 m high.

The potential series is usually truncated at a maximum degree, $n_{max}$ ($m_{max} = n_{max}$), based on accuracy requirements and computational ability. The uncertainties in parameters $\mu$, $R$ and harmonic coefficients are not significant. It is the maximum degree considered that determines the overall accuracy of the gravity model. Using a 300-degree EGM2008 model as a reference, the accuracy of models truncated at different degrees can be assessed. Fig. 2 shows the mean GGT model errors (sampled from 10°×10° grids) as functions of $n_{max}$ at different heights, where $n_{max} = 0$ and $n_{max} = 1$ represent the central and $J_2$ gravity model respectively. For altitudes between 300 km and 1000 km, the $J_2$ gravity model reduces GGT error by 2 orders of magnitude compared to the central gravity model.

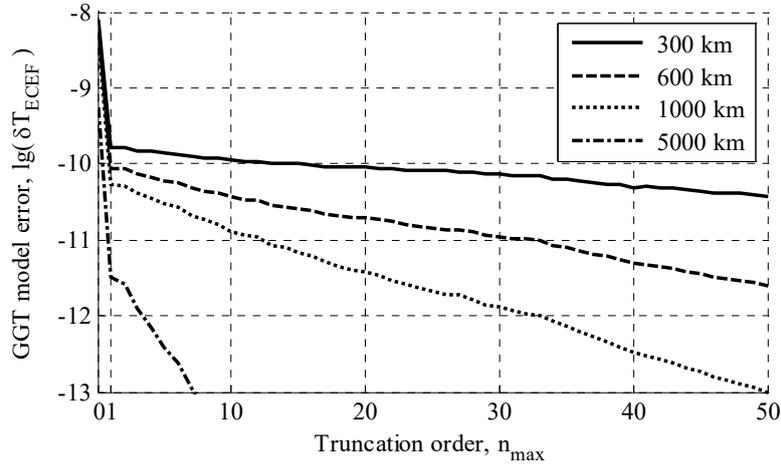

**Fig. 2 GGT model error as a function of truncation degree ($n_{max}$ = 0, 1 represent the central and $J_2$ gravity model respectively).**

Error sources of $\mathbf{T}_{ECEF}$ observations include gradiometer measurement error, attitude error and EOP error. The error propagation equation is obtained by linearization of Eq. (2):

$$\begin{aligned}\Delta \mathbf{T}_{ECEF} &= \Delta \mathbf{Q} \cdot \mathbf{W} \cdot \mathbf{T}_0 \cdot \mathbf{W}^T \cdot \mathbf{Q}^T + \mathbf{Q} \cdot \mathbf{W} \cdot \mathbf{T}_0 \cdot \mathbf{W}^T \cdot \Delta \mathbf{Q}^T \\ &+ \mathbf{Q} \cdot \Delta \mathbf{W} \cdot \mathbf{T}_0 \cdot \mathbf{W}^T \cdot \mathbf{Q}^T + \mathbf{Q} \cdot \mathbf{W} \cdot \mathbf{T}_0 \cdot \Delta \mathbf{W}^T \cdot \mathbf{Q}^T \\ &+ \mathbf{Q} \cdot \mathbf{W} \cdot \Delta \mathbf{T}_0 \cdot \mathbf{W}^T \cdot \mathbf{Q}^T\end{aligned} \qquad (9)$$

where $\Delta \mathbf{W}$ and $\Delta \mathbf{Q}$ are the rotation matrix uncertainties related to EOP error and attitude error respectively, and $\Delta \mathbf{T}_0$ is the gradiometer measurement error.

From recent geodetic reports, the relative accuracy of EOP is better than $10^{-8}$ (i.e., 1 cm with respect to 1000 km) [14]. Observation errors induced by the rotation matrix uncertainty $\Delta \mathbf{W}$ are smaller than $10^{-4}$ E and can be neglected in navigation. The attitude error mainly depends on attitude sensor noise. With current CCD star trackers, a precision of 1 arcsec can be achieved. This error causes a matrix rotation uncertainty of about $10^{-6}$, which matches the fractional precision of the gradiometer on GOCE. Table 1 lists the GGT observation errors caused by attitude errors at different altitudes.

**Table 1  GGT observation errors from attitude errors**

| Altitude, km | Attitude error, arcsec; $\mathbf{T}_{ECEF}$ observation error, E | | |
|---|---|---|---|
|  | 0.1 | 1 | 10 |
| 300 | $1.30\times10^{-3}$ | 0.0130 | 0.130 |
| 600 | $1.14\times10^{-3}$ | 0.0114 | 0.114 |
| 1000 | $9.62\times10^{-4}$ | $9.62\times10^{-3}$ | 0.0962 |
| 5000 | $2.62\times10^{-4}$ | $2.62\times10^{-3}$ | 0.0262 |

## III.  Eigen-Decomposition Method

The GGT inversion problem is to calculate the roots of a system of nonlinear equations, and can be solved by many numerical methods, one of which is the brute-force iterative technique [13]. This section investigates the eigenvalues and eigenvectors of GGTs of the central and $J_2$ gravity models and presents a semi-analytical method to solve the problem.

Recall from Eq. (2) that matrix representations of GGT in two different coordinate systems have the following relation:

$$\mathbf{T}_B = \mathbf{L}_{A\to B} \mathbf{T}_A \mathbf{L}_{A\to B}^T \tag{10}$$

where $\mathbf{T}_A$ and $\mathbf{T}_B$ are GGTs in frames A and B, and $\mathbf{L}_{A\to B}$ is the rotation matrix from A to B. The GGT matrix is diagonalizable because of symmetry. Assume that frame A is the frame in which GGT is a diagonal matrix. Then Eq. (10) becomes Eigen-Decomposition of $\mathbf{T}_B$. Frame A is called the Eigen frame and $\mathbf{T}_A$ is denoted by $\mathbf{T}_{EIG}$. The diagonal elements of $\mathbf{T}_{EIG}$ are eigenvalues and the three columns of $\mathbf{L}_{A\to B}$ are the corresponding eigenvectors.

### A.  GGTs of Central and $J_2$ Gravity Models

*1. Central Gravity Field*

Eq. (5) gives GGT in the ECEF frame. GGT in the ENU frame is:

$$\mathbf{T}_{ENU} = \begin{bmatrix} T_{EE} & T_{EN} & T_{EU} \\ T_{NE} & T_{NN} & T_{NU} \\ T_{UE} & T_{UN} & T_{UU} \end{bmatrix} = \frac{\mu}{r^3} \begin{bmatrix} -1 & 0 & 0 \\ 0 & -1 & 0 \\ 0 & 0 & 2 \end{bmatrix} \quad (11)$$

From Eq. (11), ENU is the Eigen frame and the eigenvalues are functions of only $r$. The rotation matrix from ENU to ECEF is:

$$\mathbf{L}_{ENU \to ECEF} = \begin{bmatrix} -\sin\lambda & -\sin\varphi\cos\lambda & \boxed{\cos\varphi\cos\lambda} \\ \cos\lambda & -\sin\varphi\sin\lambda & \boxed{\cos\varphi\sin\lambda} \\ 0 & \cos\varphi & \boxed{\sin\varphi} \end{bmatrix} \quad (12)$$

From Eq. (12), the eigenvectors are functions of $\varphi$ and $\lambda$. Furthermore, the eigenvector corresponding to the maximum eigenvalue is the unit vector of $\mathbf{r}$ in ECEF (as seen from the column inside the box of $\mathbf{L}_{ENU \to ECEF}$).

Let $\xi$ and $\boldsymbol{\eta}$ denote the eigenvalue and eigenvector respectively. Position $\mathbf{r}$ can be obtained by:

$$\mathbf{r} = \left(\frac{2\mu}{\xi_{max}}\right)^{\frac{1}{3}} \boldsymbol{\eta}_{max} \quad (13)$$

Where $\xi_{max}$ is the maximum eigenvalue and $\boldsymbol{\eta}_{max}$ is the corresponding eigenvector. Eq. (13) establishes the foundation of the Eigen-Decomposition method for spacecraft positioning.

*2. $J_2$ Gravity Field*

The gravitational potential of the $J_2$ gravity model truncates the harmonic summation at degree 2 and order 0:

$$U = \frac{\mu}{r}\left[1 - J_2\left(\frac{R}{r}\right)^2 \frac{1}{2}(3\sin^2\varphi - 1)\right] \quad (14)$$

where $J_2$ is the unnormalized second zonal harmonic. The elements of GGT in the ECEF frame are given by:

$$T_{xx} = \frac{\mu}{r^3}\left(-1+\frac{3x^2}{r^2}-\frac{3}{2}J_2\left(\frac{35R^2z^2x^2}{r^6}-\frac{5R^2z^2}{r^4}-\frac{5R^2x^2}{r^4}+\frac{R^2}{r^2}\right)\right)$$

$$T_{yy} = \frac{\mu}{r^3}\left(-1+\frac{3y^2}{r^2}-\frac{3}{2}J_2\left(\frac{35R^2z^2y^2}{r^6}-\frac{5R^2z^2}{r^4}-\frac{5R^2y^2}{r^4}+\frac{R^2}{r^2}\right)\right)$$

$$T_{zz} = \frac{\mu}{r^3}\left(-1+\frac{3z^2}{r^2}-\frac{3}{2}J_2\left(\frac{35R^2z^4}{r^6}-\frac{30R^2z^2}{r^4}+\frac{3R^2}{r^2}\right)\right)$$

$$T_{xy} = T_{yx} = \frac{\mu}{r^3}\left(\frac{3xy}{r^2}-\frac{3}{2}J_2\left(\frac{35R^2z^2xy}{r^6}-\frac{5R^2xy}{r^4}\right)\right) \quad (15)$$

$$T_{xz} = T_{zx} = \frac{\mu}{r^3}\left(\frac{3xz}{r^2}-\frac{3}{2}J_2\left(\frac{35R^2xz^3}{r^6}-\frac{15R^2xz}{r^4}\right)\right)$$

$$T_{yz} = T_{zy} = \frac{\mu}{r^3}\left(\frac{3yz}{r^2}-\frac{3}{2}J_2\left(\frac{35R^2yz^3}{r^6}-\frac{15R^2yz}{r^4}\right)\right)$$

GGT in the ENU frame is given by:

$$\mathbf{T}_{ENU} = \begin{bmatrix} T_{EE} & 0 & 0 \\ 0 & T_{NN} & T_{NU} \\ 0 & T_{UN} & T_{UU} \end{bmatrix} \quad (16)$$

where:

$$T_{EE} = -\frac{\mu}{r^3}+\frac{3J_2\mu R^2}{2r^5}(5\sin^2\varphi-1)$$

$$T_{NN} = -\frac{\mu}{r^3}+\frac{3J_2\mu R^2}{2r^5}(7\sin^2\varphi-3)$$

$$T_{UU} = \frac{2\mu}{r^3}-\frac{6J_2\mu R^2}{r^5}(3\sin^2\varphi-1) \quad (17)$$

$$T_{NU} = T_{UN} = \frac{6J_2\mu R^2}{r^5}\sin(2\varphi)$$

For the $J_2$ gravity field, ENU is not the Eigen frame. It is noticed that the cross terms of $\mathbf{T}_{ENU}$ involving the East direction are all zeros. Thus ENU can be transformed to the Eigen frame through rotation around the first axis (East direction). The geometric relationship of these frames is depicted in Fig. 3.

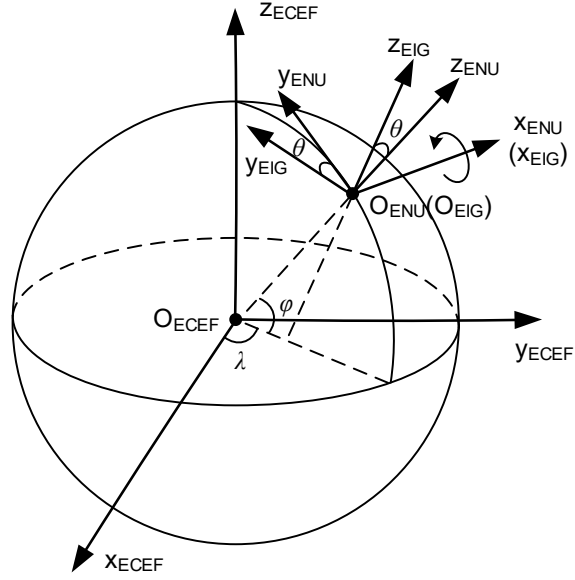

**Fig. 3 Geometric relationship between ECEF, ENU and the Eigen frame for $J_2$ gravity model.**

From Eq. (16) and Eq. (17), it can be easily proven that $T_{EE}$ is the second maximum eigenvalue, which is a function of $r$ and $\varphi$. The rotation matrix from the Eigen frame to ECEF is:

$$\mathbf{L}_{EIG \to ECEF} = \begin{bmatrix} -\sin\lambda & -\sin(\varphi+\theta)\cos\lambda & \boxed{\cos(\varphi+\theta)\cos\lambda} \\ \cos\lambda & -\sin(\varphi+\theta)\sin\lambda & \boxed{\cos(\varphi+\theta)\sin\lambda} \\ 0 & \cos(\varphi+\theta) & \boxed{\sin(\varphi+\theta)} \end{bmatrix} \quad (18)$$

where $\theta$ is the rotation angle between ENU and the Eigen frame. From Eq. (18), the eigenvector corresponding to the maximum eigenvalue is the unit vector of the z-axis of the Eigen frame (as seen from the column inside the box of $\mathbf{L}_{EIG \to ECEF}$). The longitude, $\lambda$, is independent of $\theta$. But the latitude, $\varphi$, is mixed with $\theta$.

Now consider the relationship between $\mathbf{T}_{EIG}$ and $\mathbf{T}_{ENU}$. The rotation matrix from the Eigen frame to ENU, which can be obtained by performing Eigen-Decomposition on $\mathbf{T}_{ENU}$, is:

$$\mathbf{L}_{EIG \to ENU} = \begin{bmatrix} 1 & 0 & 0 \\ 0 & \cos\theta & \boxed{\sin\theta} \\ 0 & -\sin\theta & \cos\theta \end{bmatrix} \quad (19)$$

where $\theta$ has the same sign as $\varphi$. Thus, $\theta$ can be extracted from the second component of the third eigenvector of $\mathbf{T}_{ENU}$ (the element inside the box of $\mathbf{L}_{EIG \to ENU}$).

For the $J_2$ gravity field, Eigen-Decomposition of $\mathbf{T}_{ECEF}$ yields an initial position, which can be used to calculate $\mathbf{T}_{ENU}$. Eigen-Decomposition of $\mathbf{T}_{ENU}$ then yields $\theta$, which can be used to correct the latitude $\varphi$. Meanwhile, $\varphi$ together with $\mathbf{T}_{EE}$ can be used to refine $r$.

**B. Workflow of the Algorithm**

The workflow of the Eigen-Decomposition based positioning algorithm is presented here. It is assumed that the GGT inputs of the algorithm have already been transformed to the ECEF frame.

*1. Position Initialization.*

Perform Eigen-Decomposition on $\mathbf{T}_{ECEF}$. Use the maximum eigenvalue and the corresponding eigenvector to obtain an initial position according to Eq. (13).

*2. Refinement of r.*

Calculate $\mathbf{T}_{ENU}$ according to Eq. (17) with the initial position. Refine $r$ according to the following iterative equation with $\mathbf{T}_{EE}$ and $\varphi$:

$$r = \left[ -\mu \bigg/ \left( \mathbf{T}_{EE} - \frac{3J_2 \mu R^2}{2r_0^5}\left(5\sin^2\varphi - 1\right) \right) \right]^{1/3} \qquad (20)$$

*3. Correction for $\varphi$.*

Use the refined $r$ to update $\mathbf{T}_{ENU}$. Perform Eigen-Decomposition on $\mathbf{T}_{ENU}$. Then $\theta$ can be obtained from the third eigenvector. Correct $\varphi$ with $\theta$ as follows:

$$\varphi = \varphi_0 - \theta \qquad (21)$$

*4. Repeat 2 and 3.*

Repeat 2 and 3 until the numerical accuracy requirement is met. The final position can be obtained using the refined $r$, the corrected $\varphi$, and the initial $\lambda$.

## IV. Simulation and Real Data Tests

**A. GGT Inversion Accuracy**

5°×5° gridded GGTs at four different altitudes are simulated to test the Eigen-Decomposition algorithm at the global scale. The numerical precision of the algorithm is investigated using $\mathbf{T}_{ECEF}$ computed from the $J_2$ gravity model without considering any noise. Statistical three-dimensional (3D) position errors are summarized in Table 2.

The algorithm implemented here repeats *step 2* and *step 3* only once, and the maximum errors are all below 0.1 m. It is noticed that the numerical solutions are more accurate at higher altitudes. That is because the $J_2$ effect becomes weaker at higher altitudes, and thus the initial positions can be more easily corrected.

**Table 2  Numerical precision of the Eigen-Decomposition algorithm**

| Altitude, km | Statistical 3D position error, m | | |
|---|---|---|---|
|  | Maximum | Minimum | Mean |
| 300 | 0.0885 | $9.93\times10^{-5}$ | 0.0525 |
| 600 | 0.0713 | $7.30\times10^{-5}$ | 0.0421 |
| 1000 | 0.0539 | $4.94\times10^{-5}$ | 0.0319 |
| 5000 | $6.14\times10^{-3}$ | $2.37\times10^{-6}$ | $3.66\times10^{-3}$ |

To simulate noisy environments, the 300-degree EGM2008 gravity model is used to generate $\mathbf{T}_{ECEF}$ as true values and errors at levels of 1 E, 0.1 E, 0.01 E, and 0.001 E are added. The mean 3D position errors are given in Table 3. The results are consistent with error analysis given in Eq. (7). For example, the mean GGT model error of the $J_2$ gravity model at the altitude of 300 km is 0.163 E, and the vertical and horizontal sensitivity factors are $6.0\times10^{-3}$ E/m and $1.2\times10^{-4}$ E/m respectively. With an additional observation error of 0.1 E, the vertical and horizontal position errors will be 432 m and 216 m, which are consistent with the mean position error of 421 m in Table 3.

From Table 3, it is shown that at a given altitude the position accuracy improves with decreases in observation error. However, accuracy improvement slows down when the observation error is smaller than the $J_2$ model error. In addition, a limit exists when the observation error gets closer to zero. For the case of 300 km altitude, the position accuracy limit is 326 m. For the case of 5000 km altitude, the limit is 59.5 m.

**Table 3  GGT inversion accuracy in noisy environments**

| Altitude, km | $\mathbf{T}_{ECEF}$ observation error, E; Mean 3D position error, m | | | |
|---|---|---|---|---|
|  | 1 | 0.1 | 0.01 | 0.001 |
| 300 | $2.69\times10^{3}$ | 421 | 328 | 326 |
| 600 | $3.19\times10^{3}$ | 388 | 224 | 221 |
| 1000 | $3.95\times10^{3}$ | 431 | 174 | 169 |
| 5000 | $2.28\times10^{4}$ | $2.23\times10^{3}$ | 231 | 59.5 |

Another significant characteristic is the relationship of the altitude. Given an observation error, position accuracy firstly increases until an altitude where the $J_2$ model error is smaller than the observation error. Above this critical altitude, the position error increases due to lower sensitivity. The critical altitude for cases of 0.1 E and 0.01 E errors are 600 km and 2000 km, and the corresponding mean position errors are 388 m and 131 m.

The position error distribution with longitude and latitude is also investigated. Fig. 4 shows the case of altitude of 600 km and observation error of 0.01 E. Because the $J_2$ model error dominates the observation error, the position error is greatly correlated with the $J_2$ model error, which is shown in Fig. 5. The position accuracy above oceans is usually better than that above continents. That is because the $J_2$ model fits well above the ocean areas.

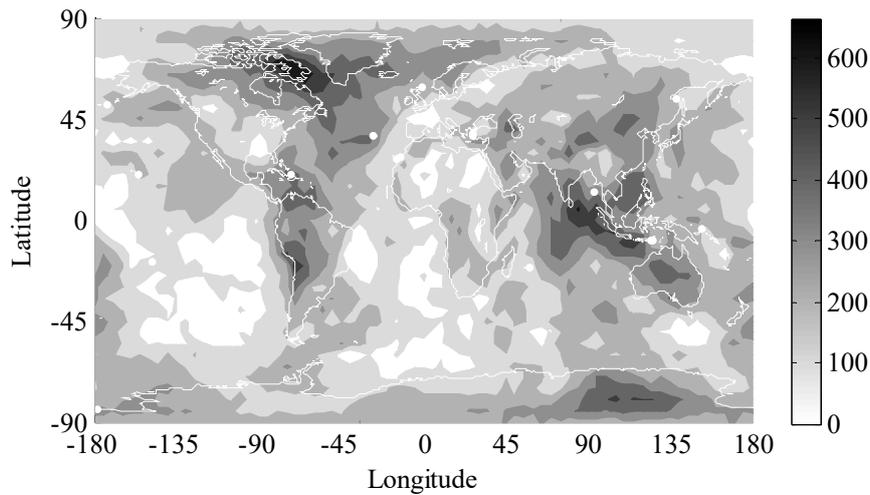

**Fig. 4  3D Position errors (m) at the altitude of 600 km with 0.01 E observation error, interval of 100.**

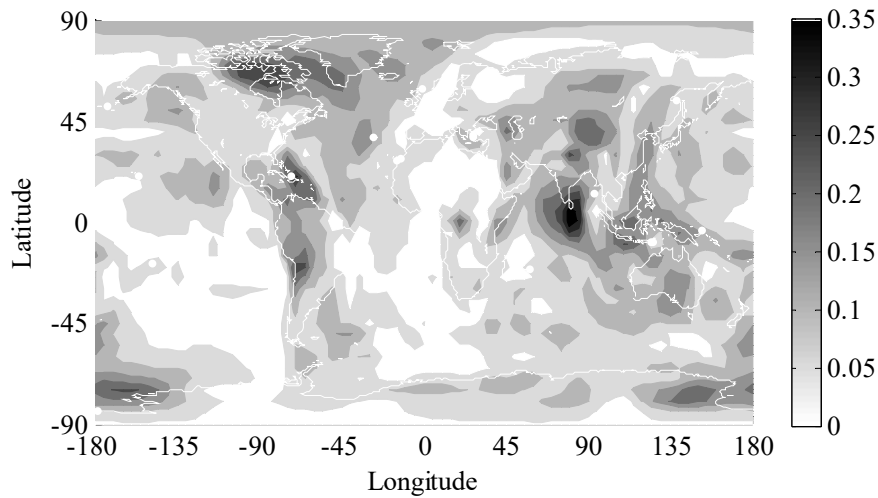

**Fig. 5  $J_2$ gravity model error (E) at the altitude of 600 km, interval of 0.05.**

**B. Performance for Spacecraft Navigation**

A 24-hour LEO orbit trajectory is simulated considering perturbations including non-spherical gravity, solar and lunar gravity, and atmospheric drag. The orbit is nearly circular, having a height of 300 km and an inclination of 80 deg. GGT measurements are generated using the 300-degree EGM2008 gravity model at a rate of 0.1 Hz, and white noise with a standard deviation of 0.1 E is added to the signals. The gradiometer reference frame is assumed to constantly coincide with the Local Vertical Local Horizontal frame (i.e., the X axis is outward along the radial, the Z axis is in the direction of the cross product of the position and velocity, and the Y axis is in the direction of motion). Attitudes measurements are provided in the form of Euler angles, to which white noise with a standard deviation of 1 arcsec is added.

As shown in Fig. 6, the accuracy of the position solutions is on the order of hundreds of meters with a mean error of 433 m. From Table 1, an attitude error of 1 arcsec at a height of 300 km causes a $\mathbf{T}_{ECEF}$ observations error of 0.0130 E. Taking the gradiometer noise 0.1 E into account, the total $\mathbf{T}_{ECEF}$ observation error is 0.113 E. Referring to the case of 300 km altitude and 0.1 E observation error in Table 3, the mean position error is 421 m, which is consistent with the mean error of position solutions.

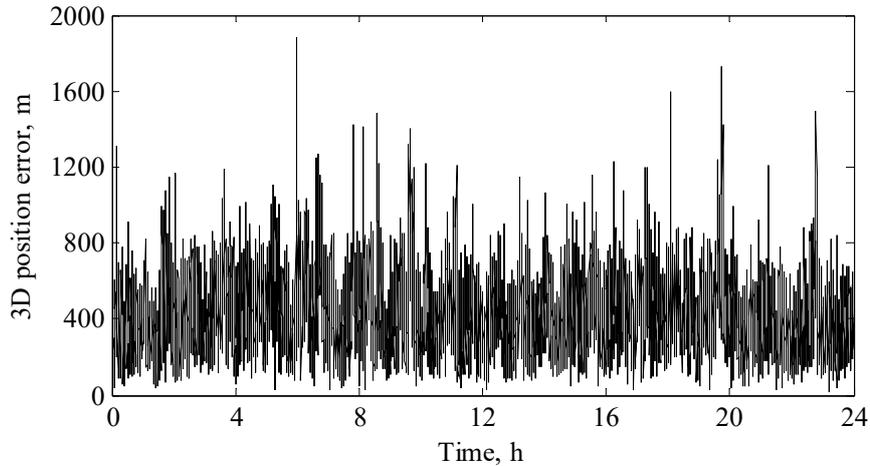

**Fig. 6  3D Position errors of the simulated LEO satellite.**

To reduce noise and improve position accuracy, a least-squares batch estimator is used. Positions are smoothed by a $J_2$ orbital dynamics model, and the velocities are estimated simultaneously. The batch interval equals one orbital period. The 3D errors of the smoothed positions and velocities are shown in Fig. 7. The mean position error is reduced to 121 m, and the mean velocity error is 0.119 m/s.

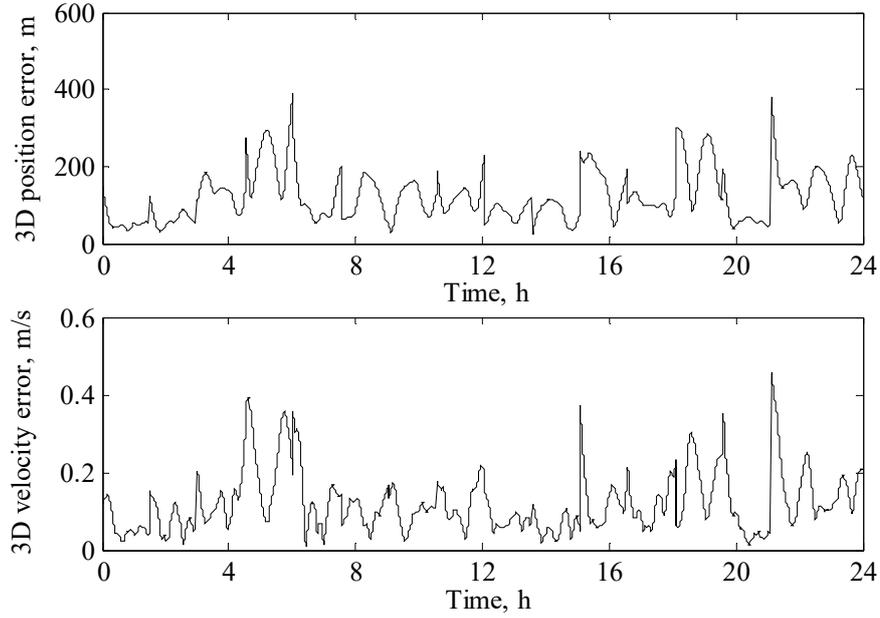

**Fig. 7 3D Position and velocity errors of the simulated LEO satellite after orbital dynamic smoothing.**

The GGT inversion navigation system is also tested on real data from the GOCE satellite [15]. In-flight calibrated gravity gradients collected on 8 September 2013 as well as the post-processed accurate attitude quaternions are used as inputs, and the resolved positions are compared with the Precise Science Orbit solutions (PSO, 2 cm RMS) derived from the GPS measurements.

The GOCE GGT measurements are provided at a rate of 1 Hz. Within measurement bandwidth (MBW) between 5 mHz and 0.1 Hz, the components $T_{xx}$, $T_{yy}$, $T_{zz}$, and $T_{xz}$ have similar noise density levels of about 0.011 E/√Hz, whereas the noise density levels of $T_{xy}$ and $T_{yz}$ are much higher, 0.35 E/√Hz and 0.50 E/√Hz respectively. Below MBW the measurement error power spectral density has a 1/$f$ behaviour. A Fourier compensation model is used here to remove the 1/$f$ error before using the measurements for test:

$$\gamma = \Delta\gamma + \gamma' \cdot t + \sum_{k=1}^{K}\left[a_k \cos 2\pi k f_0 t + b_k \sin 2\pi k f_0 t\right] \tag{22}$$

where $\gamma$ represents the 1/$f$ error, $\Delta\gamma$ is bias, $\gamma'$ is trend, $f_0$ is the base frequency and equals the orbit revolution frequency, and $a_k$ and $b_k$ are Fourier coefficients. The maximum order of the sinusoid terms $K$ is chosen to satisfy that $Kf_0 = 5$ mHz.

The 3D errors of the position solutions are shown in Fig. 8. The mean error is 838 m. The attitude quaternions were determined from a combination of star sensor and gradiometer data and have aprecision of a few arcsec [16].

From Table 1, the attitude error causes a GGT transformation uncertainty of 0.013-0.130 E for GOCE near the altitude of 300 km. With the gradiometer noise of 0.5 E added, the total observation error is about 0.513-0.63 E. Referring to Table 3, the expected position accuracy is between 421 m and 2690 m, which is consistent with theaccuracy of the results.

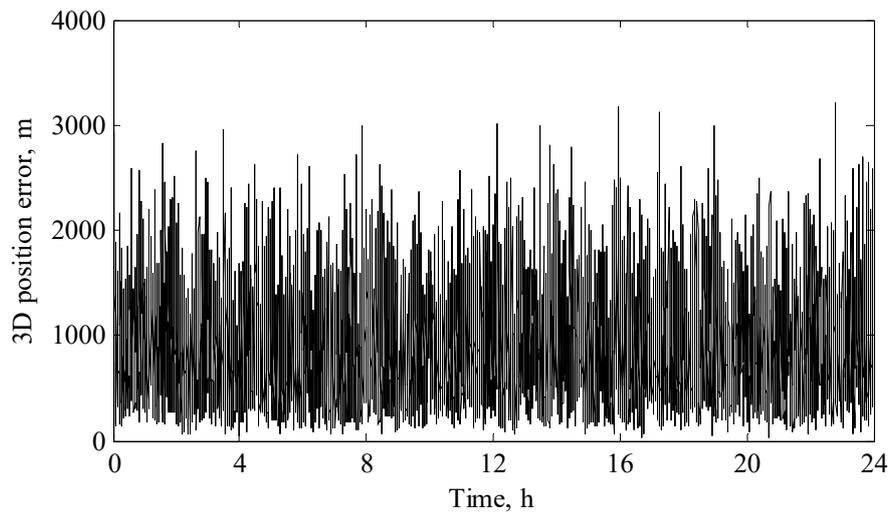

**Fig. 8  3D Position errors of GOCE varying with time.**

The same least-squares smoothing procedure is applied to GOCE. The 3D errors of the smoothed positions and velocities are shown in Fig. 9. The mean position error is reduced to 120 m, and the mean velocity error is 0.125 m/s.

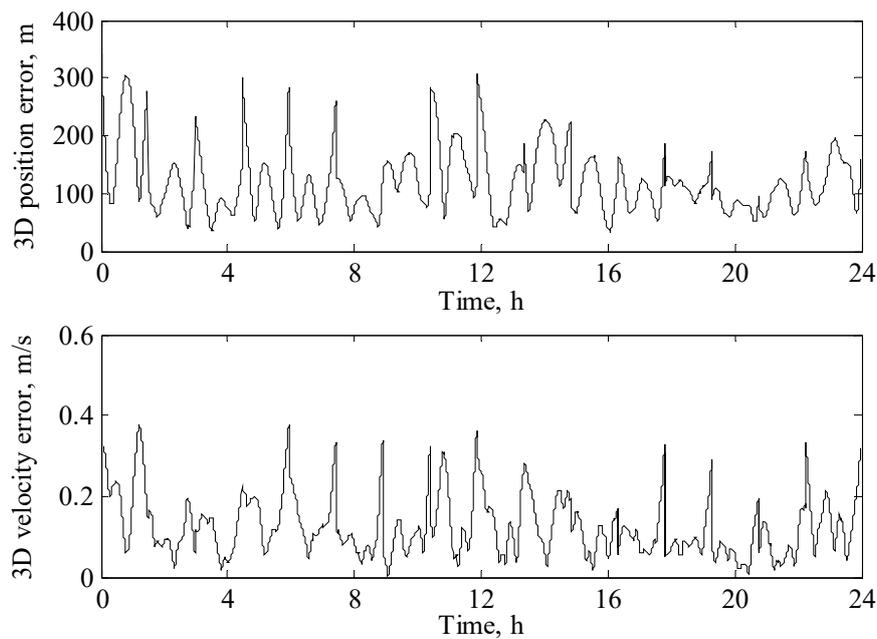

Fig. 9  3D Position and velocity errors of GOCE after dynamic smoothing.

## V.     Conclusion

In this Note, an Eigen-Decomposition algorithm for spacecraft positioning using gravity gradient measurements is presented. Position accuracy of hundreds of meters is demonstrated for low Earth orbiting satellites based on real flight data from GOCE. The proposed navigation system is suitable for spacecraft around the Earth as well as planetary bodies of which the gravity fields are known.

## Acknowledgments

This research was supported by the National Natural Science Foundation of China (No. 11002008). The authors kindly acknowledge the provision of the GOCE data by ESA. The authors also appreciate Christophe Macabiau and anonymous reviewers' constructive comments and suggestions.